\documentclass[aps,prl,amsmath,amssymb,superscriptaddress,showkeys,reprint]{revtex4-1}
\usepackage{graphicx}
\usepackage{dcolumn}
\usepackage{bm}

\begin{document}

\title{Linear Laser Tuning Using a Pressure-Sensitive
Microbubble Resonator}

\author{Ramgopal Madugani}
\affiliation{Light-Matter Interactions Unit, Okinawa Institute of Science and Technology Graduate University, Onna, Okinawa 904-0495, Japan}
\affiliation{Physics Department, University College Cork, Cork, Ireland}
\author{Yong Yang}
\affiliation{Light-Matter Interactions Unit, Okinawa Institute of Science and Technology Graduate University, Onna, Okinawa 904-0495, Japan}
\affiliation{National Engineering Laboratory for Fiber Optics Sensing Technology, Wuhan University of Technology, Wuhan, 430070, China}
\author{Vu H Le}
\affiliation{Light-Matter Interactions Unit, Okinawa Institute of Science and Technology Graduate University, Onna, Okinawa 904-0495, Japan}
\affiliation{Now at Institute of Microstructure Technology, Karlsruhe Institute of Technology, Germany}
\author{Jonathan M. Ward}
\affiliation{Light-Matter Interactions Unit, Okinawa Institute of Science and Technology Graduate University, Onna, Okinawa 904-0495, Japan}
\author{S\'{\i}le {Nic Chormaic}}
\email{sile.nicchormaic@oist.jp}
\affiliation{Light-Matter Interactions Unit, Okinawa Institute of Science and Technology Graduate University, Onna, Okinawa 904-0495, Japan}

\begin{abstract}
The tunability of an optical cavity is an essential requirement for many areas of research.  Here, we use the Pound-Drever-Hall technique to lock a laser to a whispering gallery mode (WGM) of a microbubble resonator, to show that linear tuning of the WGM, and the corresponding locked laser, display almost zero hysteresis. By applying aerostatic pressure to the interior surface of the microbubble resonator, optical mode shift rates of around $58$ GHz/MPa are achieved. The microbubble can measure pressure with a detection limit of $2\times 10^{-4}$ MPa, which is an improvement made on pressure sensing using this device. The long-term frequency stability of this tuning method for different input pressures is measured. The frequency noise of the WGM measured over $10$ minutes for an input pressure of $0.5$ MPa, has a maximum standard deviation of $36$ MHz.
\end{abstract}

\maketitle

\section{Introduction}
Whispering gallery mode (WGM) resonators are widely used across many research fields such as cavity quantum electrodynamics (cQED) \cite{Aoki2006}, nonlinear optics \cite{Yuta2015} and sensing \cite{Vollmer2008,WardLiqCoreTsense2013,YangQuasi-droplet2014}. In many applications, the WGM is required to be tunable. In cQED systems, for example, the strength of the interaction between atoms and  cavity photons relies on the cavity quality ($Q$) factor and the optical mode volume ($V$) as $Q/V$. Therefore, whispering gallery resonators (WGRs) are excellent candidates for such studies 
since they can have an ultrahigh $Q$-factor. However, there is a practical requirement in that the cavity mode should be in resonance with the atomic transitions of interest and the only realistic way to achieve this is via cavity mode tuning.

To date, several tuning techniques in WGM resonators have been demonstrated. The most widely used is temperature tuning \cite{Aoki2006,ward2010thermo,watkins2012thermo}; however, the total tuning range achievable is limited and the process is relatively slow. Additionally, in order to achieve repeatability, accurate temperature control is required, thereby increasing the complexity of the experimental setup. For a larger tuning range, tunable microresonators based on strain and stress were developed using double stem microspheres \cite{madugani2012terahertz} and bottle microresonators \cite{Pollinger2009}. However, the observed mode shift due to the mechanical displacement caused by stretching (or compressing) the WGRs with piezo stacks is also usually nonlinear \cite{madugani2012terahertz}, since a hysteresis effect is introduced when a high voltage is applied to the piezo. Other techniques that rely on etching \cite{white2005tuning}, the optical gradient force  \cite{Wiederhecker2009}, and electric \cite{ioppolo2009tuning} or magnetic \cite{ioppolo2010magnetorheological} effects have also been used, but they provide a very narrow tuning range, poor precision/repeatability, and/or an increased footprint.

More recently, an alternative WGR geometry that can support high $Q$ WGMs was developed, known as the microbubble \cite{Sumetsky:10,Watkins:11}.  It shows promise in sensing \cite{YangQuasi-droplet2014,li2010analysis} and demonstrates good tunability through several means, such as stretching \cite{Sumetsky:10} or via internal aerostatic pressure \cite{Henze:11}.  Previously, we have achieved a sensitivity of 380 GHz/MPa by decreasing the microbubble wall thickness \cite{Yang:2016}. Such tunability has already been used for applications in coupled-mode induced transparency with a single WGR \cite{Yang2015}. However, no work has been reported on the repeatability, long term stability, and frequency noise measurements of WGMs with aerostatic pressure tuning.
\section{Experimental details}
To measure the long term stability, in the conventional way, a tunable probe laser is scanned while continuously monitoring the frequency position of the WGMs. There are several factors influencing the WGM resonance frequency, such as thermal instability due to laser heating and coupling gap fluctuations caused by environmental factors. To avoid unwanted jittering between the WGM and the laser, the Pound-Drever-Hall (PDH) frequency locking technique \cite{Swaim:13} is generally employed, where the laser source is tuned into resonance with the WGM to be tested and locked to the bottom of the resonance dip \cite{Carmon:05}. The error signal from the lock-in loop can be used to monitor the WGM jittering noise. This method has also been used to counteract the noise from the noncontact coupling gap between a tapered fiber and a bottle resonator \cite{Junge2011}.  In some cases, the WGM resonator can be used as a feedback element to the laser, leading to a narrowing of the laser linewidth \cite{Liang2010} and/or improved stability of the laser source \cite{Fescenko2012}. Similarly,  locking  tunable lasers to a strain-tunable, fused silica microresonator \cite{Rezac:01} has been demonstrated. Although the tunability in this case was large, it was at the expense of hysteresis due to the nonlinearity of the piezo stacks providing the strain.

To highlight the advantages of the pressure tuning method, a microbubble was fabricated from a microcapillary using two counterpropagating CO$_2$ laser beams. The outer and inner diameters of the capillary were 350 $\mu$m and 250 $\mu$m, respectively. Using the CO$_2$ laser the capillary was initially tapered down to a diameter of 33 $\mu$m. Then, compressed air was sent through the microcapillary and the CO$_2$ beams were reapplied to form a spherical shape in the area where the CO$_2$ beams were focused. A microbubble with a diameter of 100 $\mu$m, as measured under a microscope, was selected for this experiment. The wall thickness was estimated to be around 1.4 $\mu$m \cite{Henze:11}. Laser light at 1.55 $\mu$m was coupled into the microbubble via the evanescent field of a tapered optical fiber. To eliminate coupling gap noise, the taper was placed in contact with the microbubble. The laser was scanned over 35 GHz and the WGM spectrum was recorded from the transmitted power at the output of the tapered fiber. To isolate the setup the experiment was put in an enclosure. To reduce thermal effects a low input laser power of around 10 $\mu$W was used. A high $Q$ mode with a linewidth of 112 MHz ($Q=1.6\times10^6$) was used for locking using the PDH method. 

\begin{figure} [htb]
\centering
\includegraphics[width=9cm]{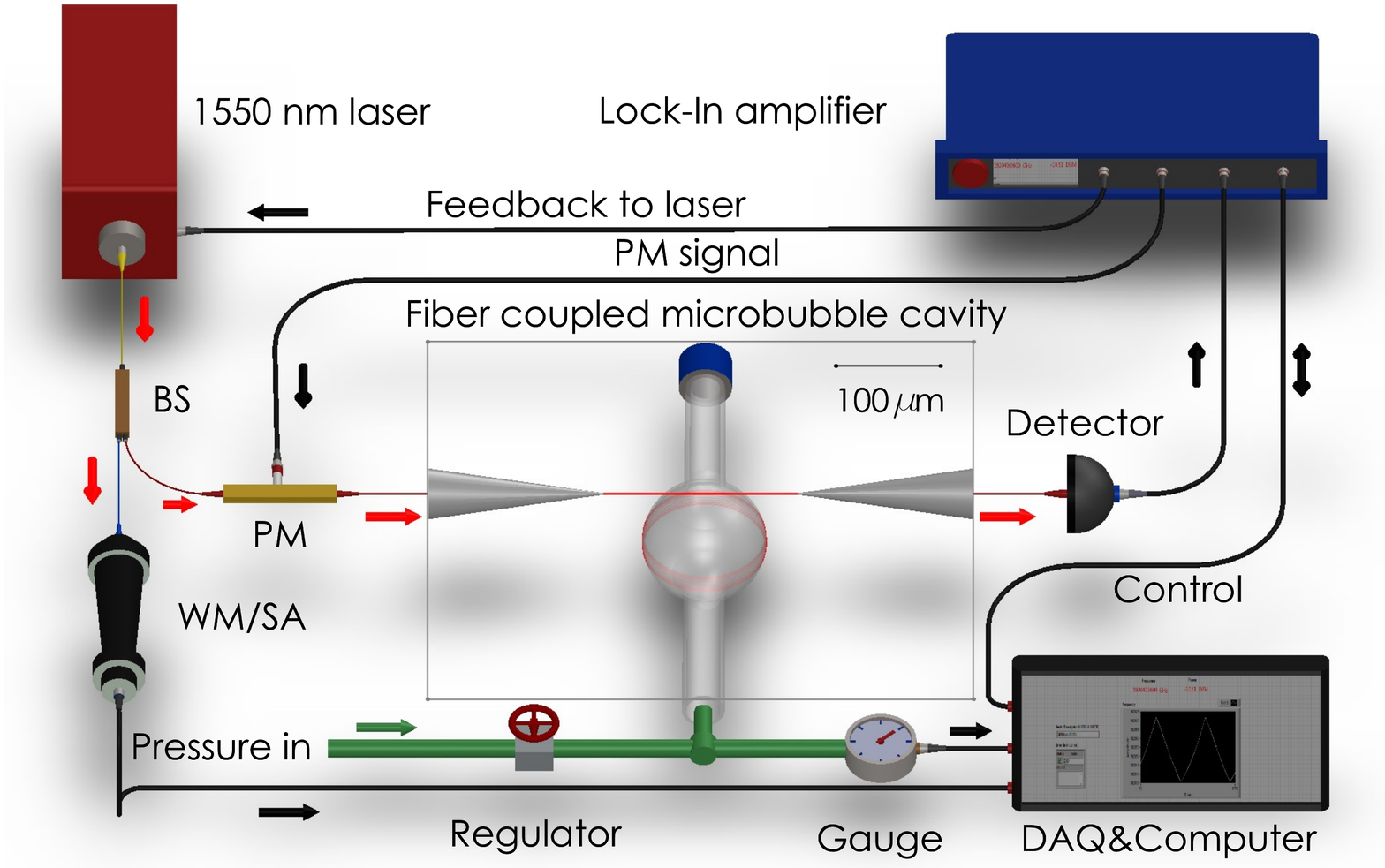}
\caption{Experimental setup. A fiber-coupled 1550 nm laser beam is  beam split (BS), with one part  sent to a spectrum analyzer (SA) and the other, with 41 MHz phase modulation (PM), to the fiber taper to couple with the microbubble (Inset). The transmission signal was sent to the laser feedback loop control. Once the laser locks to the WGM, the bubble is internally pressurized with nitrogen gas while monitoring the lock-in amplitude noise.  The pressure gauge and wavemeter (WM) (or spectrum analyzer) signals were simultaneously recorded through a LabVIEW interface.}
\label{scheme}
\end{figure}

Fig. \ref{scheme} shows the scheme of the PDH locking system. The light from the tunable laser source (Newfocus Velocity 6700) was initially split by an inline beam splitter. One of the splitter outputs was used in the feedback loop for the PDH locking. The light launched into the taper was phase modulated at 41 MHz by an electro-optical modulator. The transmitted light after the microbubble-taper system was detected on a photodiode. The photodiode signal was mixed with the driving signal in the lock-in amplifier (HF2LI model Zurich Instruments) for demodulating the error signal, which was sent to the built-in PID controller. Using the PID parameters obtained by the Ziegler-Nichols method, a feedback signal was generated and sent back to the piezo controller of the laser source allowing the laser frequency to follow any WGM shift. The frequency of the locked laser output was monitored by connecting the second beam splitter output to a laser frequency spectrum analyzer.
\section{Results and discussion}
\subsection{Hysteresis measurements}
\begin{figure}[htb]
\centering
\includegraphics[width=8.5cm]{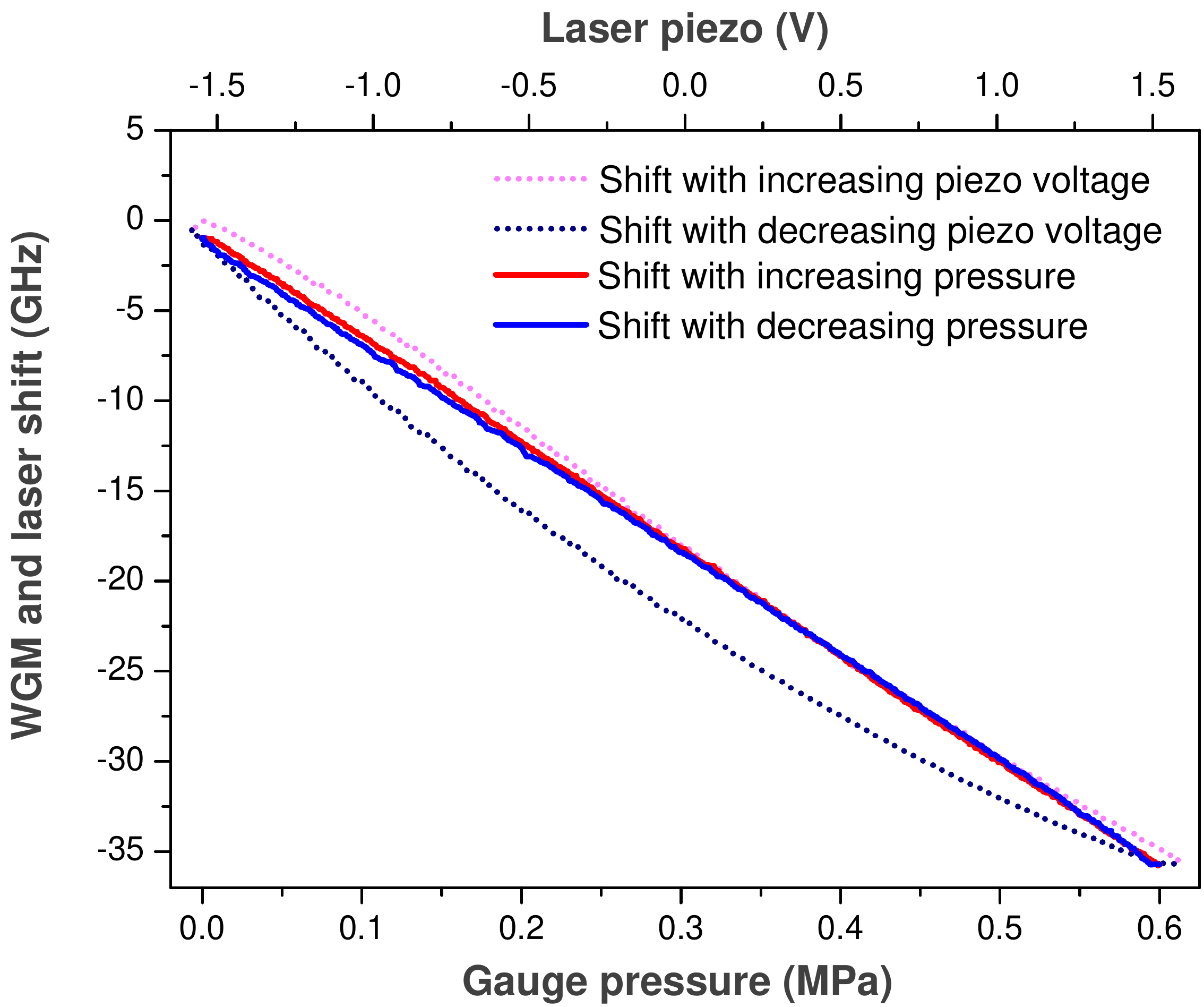}
\caption{Hysteresis measurements: Laser frequency shift as a function of applied pressure. Solid red (blue)  curve represents the laser frequency change for applied pressure in increasing (decreasing) direction. For comparison the hysteresis curve of the unlocked laser frequency shift as a function of applied piezo voltage is shown in dotted magenta for increasing and dark blue for decreasing piezo voltage directions.}
\label{Hysteresismeasurement}
\end{figure}

For pressure tuning, the input of the microbubble was connected to a nitrogen gas source, adjustable by a manual pressure regulator. The input pressure was measured using an electronic pressure sensor. 
Using the measured dimensions of the bubble the sensitivity of the WGM to input pressure was calculated as $61.4$ GHz/MPa \cite{Yang:2016}. With 0.6 MPa gauge pressure, the frequency of the WGM can be tuned by 34.8 GHz, which is the maximum fine tuning range accessible for the laser in use. A wavemeter was used for measuring the frequency of the laser. Fig. \ref{Hysteresismeasurement} is a plot of the laser frequency shift when the pressure was increased from 0 MPa to 0.6 MPa and back to 0 MPa.
The measured sensitivity is $58$ GHz/MPa. The red curve represents  increasing pressure. After the microbubble reached the high pressure value, the pressure was slowly decreased back to see whether the laser frequency returned to the same position with the same pressure value, illustrated in Fig. \ref{Hysteresismeasurement} as the blue curve. Note, the wavemeter has a resolution of around 100 MHz and, within this resolution, the tuning curve is almost linear without much hysteresis except in the low pressure range where up to 700 MHz hysteresis was seen. In comparison, the free running or unlocked laser tuning (via the laser's internal piezo drive) profile has a maximum hysteresis of 5 GHz, also shown in Fig. \ref{Hysteresismeasurement}. Hence, when the laser is locked to the pressure tunable microbubble it has improved linearity and almost no hysteresis.

\subsection{Long term stability}
\begin{figure}[htb]
\centering
\includegraphics[width=8.5cm]{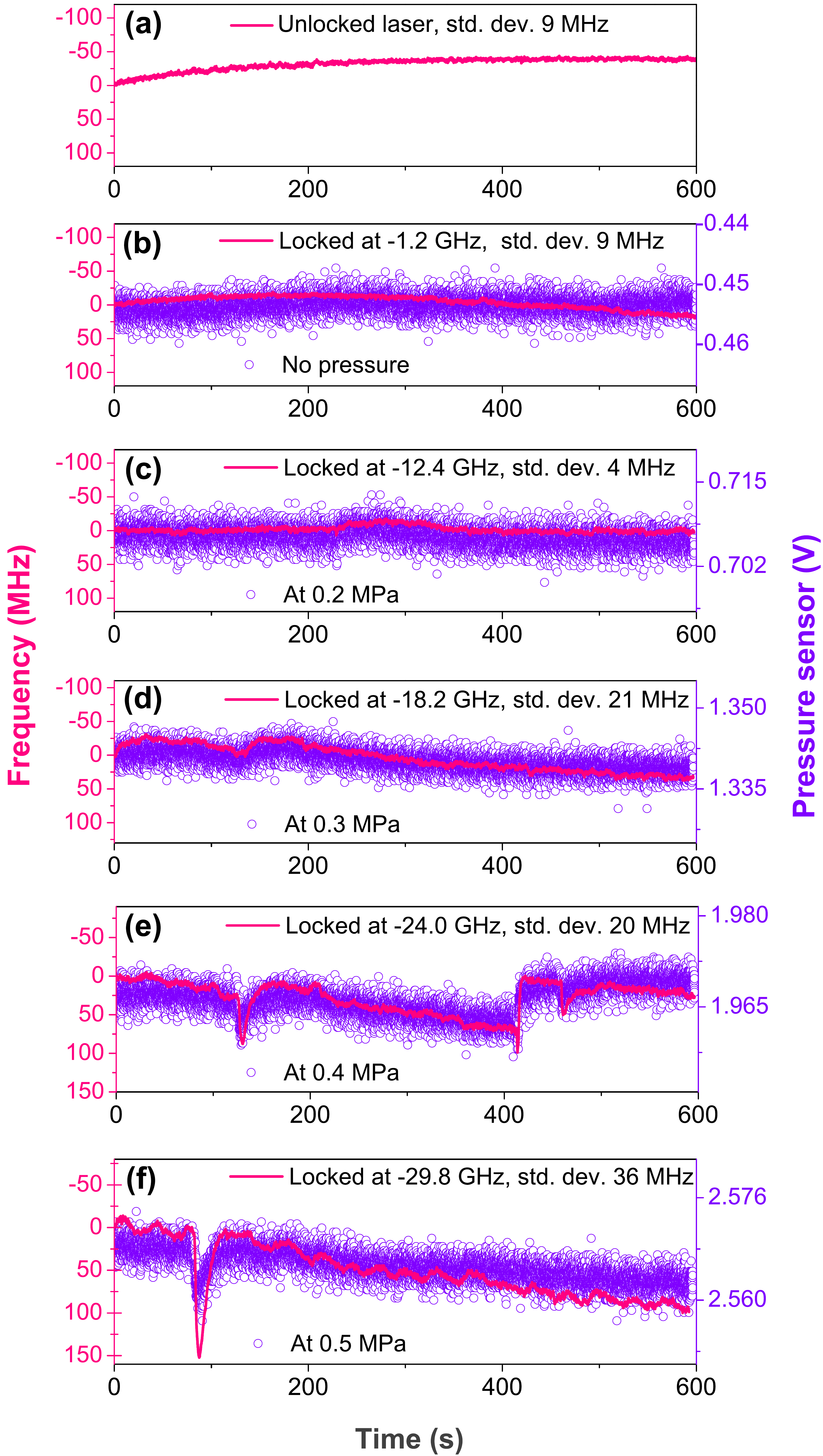}
\caption{Long term stability: Frequency as a function of time for (a) the free running laser with no lock  and the microbubble WGM locked laser at (b) no pressure, (c) 0.2 MPa (d) 0.3 MPa (e) 0.4 MPa and (f) 0.5 MPa. Each pressure increment corresponds to locking the laser to the WGM at different points along the linear plot in Fig. \ref{Hysteresismeasurement}.}
\label{Fig:Longtermstability}
\end{figure}

For an optical system containing a tunable element, it is important to know the long term stability. Here, we have shown a comparison of such stability for different input pressures. First, the laser is left unlocked for ten minutes, see Fig. \ref{Fig:Longtermstability}(a), to determine the free running frequency noise of the laser. In the rest of the cases, from Figs. \ref{Fig:Longtermstability}(b) to 3(f), the pressure was fixed at values ranging from 0 MPa to 0.5 MPa and, at each pressure value, both the pressure sensor voltage and frequency fluctuations were monitored for more than 10 min. To get a better frequency resolution, the wavemeter was replaced by a Thorlabs SA200 laser spectrum analyzer (SA) with a free spectral range (FSR) of 1.5 GHz and a resolution of 2 MHz. The Fabry-Perot cavity of the SA was scanned and the resulting mode was monitored for the locked laser frequency fluctuations. Standard deviation values of the frequency stability are given in Fig. \ref{Fig:Longtermstability} for each value of applied pressure.

During the 10 minutes observation time, the laser frequency follows the jittering of the WGM. With no input pressure the measured voltage signal from the pressure sensor shows a background noise level of $\sigma_P$ = $1.43\times 10^{-3}$ V. The electronic pressure sensor has a rated sensitivity, $S = 12.3$ V/MPa, therefore the limit of detection (LOD) = (3$\sigma_P$ / S) = $3.5\times 10^{-4}$ MPa, which is equivalent to a frequency LOD of 19.7 MHz. In comparison, the LOD of the WGM microbubble sensor to pressure is determined by its sensitivity and resolution, which depend on the optical linewidth. From the measurements, the locked laser frequency shows a variation of $\sigma_L$ = 9.5 MHz, hence equivalently this microbubble should have a LOD of $1.6\times 10^{-4}$ MPa. However, the limit of detection of the overall measurement system is a combination of the linewidth of the spectrum analyzer (2 MHz) and the microbubble sensitivity, yielding a significantly improved LOD of $3.4\times 10^{-5}$ MPa. Therefore, the electronic pressure sensor has a much poorer LOD, so there may be pressure changes that are too small for it to detect.  This is clearly evident in Fig. \ref{Fig:Longtermstability}(f). In Fig. \ref{Fig:Longtermstability}, the long term trend indicates that the locked laser follows the measured pressure sensor signal drift, with some random jumps at higher pressures.  These jumps are suspected to be due to instability of the pressure source or regulator. Above all, excluding the obvious pressure changes, the frequency fluctuations of the WGM are within a standard deviation of 36 MHz at any pressure. Note that the WGM linewidth was 112 MHz and the system was left free running without compensation of other factors such as temperature.  The long term stability seems satisfactory for practical usage.

\subsection{Noise spectrum}
\begin{figure}[htb]
\centering\includegraphics[width=8.5cm]{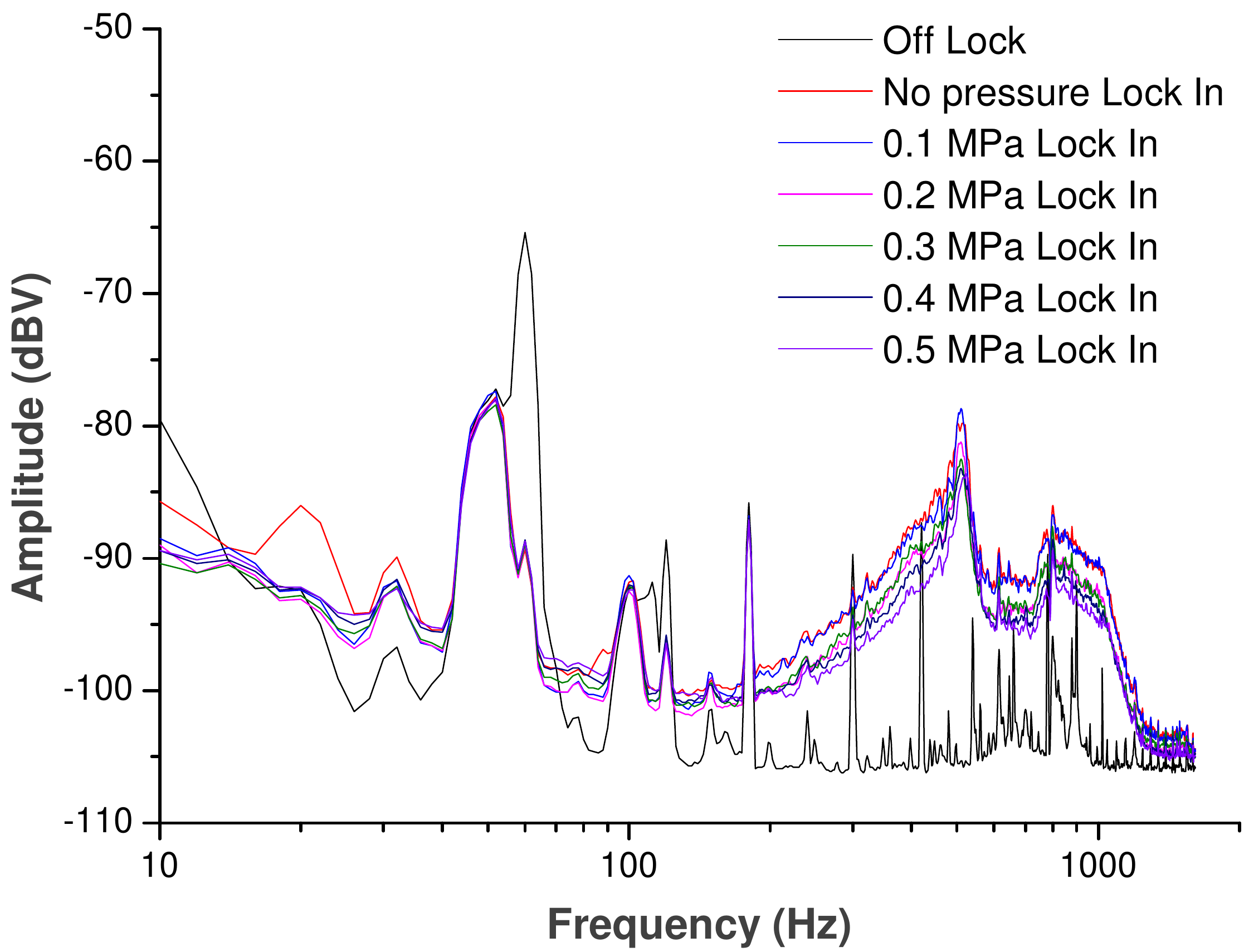}
\caption{Low frequency amplitude noise profiles of the locked laser from zero applied pressure to 0.5 MPa applied pressure, as measured on the gauge, and a comparison to its unlocked state (black curve).}
\label{Fig:Lockingcomparison}
\end{figure}

The amplitude noise of the laser output was also measured with a Fourier transform spectrometer, see Fig. \ref{Fig:Lockingcomparison}. The low frequency noise spectrum of the laser from 10 Hz to 1.6 kHz was recorded when the laser was locked to the WGM for pressures from 0 to 0.5 MPa. For comparison, a spectrum was also taken when the laser was unlocked. Firstly, the microbubble resonator with no input pressure is inherently more noisy than the laser. Secondly, and most importantly for the pressure tuning method, the spectra show that the addition of pressure does not significantly increase the low frequency noise of the laser. In fact, the overall noise level decreases with increasing pressure. As evident from Fig. \ref{Fig:Lockingcomparison}, a drop in noise for an input pressure of 0.5 MPa compared to the 0.1 MPa pressure input was observed. This also confirms the observation in the hysteresis curve where a larger hysteresis is observed for lower pressure levels.

The ratio of the phase modulation frequency and the WGM half linewidth is 0.7 and defines the strength of locking \cite{Bjorklund1983}. Accordingly, the locking achieved is weak and leads to the apparent noise in Fig. \ref{Fig:Lockingcomparison}. To reduce the noise, a better lock is required. In this case, one should either increase the phase modulation frequency or use a microbubble with a higher \textit{Q}, both of which are technically possible for further development.

\section{Conclusion}
In summary, we experimentally demonstrated that  aerostatic pressure tuning of a microbubble is a good alternative to the WGR tuning methods. This method has the advantages of linear tuning with no hysteresis, long-term stability, and a simple experimental configuration. As a prospective advantage, in the coupled resonators scheme with multiple resonators \cite{Xiao2007,dumeige2009stopping}, microbubbles can be locally tuned by connecting them to separate pressure sources, whereas this would be technically challenging for temperature tuning or other methods. Our experiment also realized, in principle, an LOD improvement for aerostatic pressure sensing in a microbubble. 

\section*{Acknowledgments}
This work was supported by Okinawa Institute of Science and Technology Graduate University.


\end{document}